\documentclass[prx,twocolumn,superscriptaddress,nofootinbib,longbibliography]{revtex4-2}
\usepackage{graphicx}
\usepackage{bm}
\usepackage{amssymb}
\usepackage{color}
\usepackage{amsmath}
\usepackage{mathrsfs}
\usepackage{amstext}
\usepackage[english]{babel}
\usepackage{latexsym}
\usepackage{array}
\usepackage{ulem}
\usepackage{multirow}         
\usepackage[usenames,dvipsnames]{xcolor}
\usepackage[colorlinks=true,citecolor=Blue,linkcolor=RubineRed,urlcolor=Blue]{hyperref}
\usepackage{physics}
\usepackage{enumerate}

\newcommand{\expv}[1]{\langle #1 \rangle}

\newcommand{\pap}[1]{\left( #1 \right)}
\newcommand{\pas}[1]{\left[#1 \right]}

\begin{document}
\title{From edge to bulk: Cavity induced displacement of topological non-local qubits}
\author{F. P. M. M\'endez-C\'ordoba}
\email{fp.mendez10@uniandes.edu.co}
\affiliation{Departamento de F\'isica, Universidad de Los Andes, A.A. 4976, Bogot\'a, Colombia}

\author{F. J. Rodr\'iguez}
\affiliation{Departamento de F\'isica, Universidad de Los Andes, A.A. 4976, Bogot\'a, Colombia}

\author{C. Tejedor}
\affiliation{Departamento de F{\'i}sica Te{\'o}rica de la Materia Condensada and Condensed Matter Physics Center (IFIMAC), Universidad Aut{\'o}noma de Madrid, Madrid 28049, Spain}

\author{ L. Quiroga}
\affiliation{Departamento de F\'isica, Universidad de Los Andes, A.A. 4976, Bogot\'a, Colombia}
\date{\today}
\begin{abstract}

We investigate the ability of selective cavity coupling to a topological chain for tailoring the connectivity of Majorana fermions. We show how topological qubits (TQs), associated with non-local Majorana fermion pairing, can be displaced from the edges to the bulk of a topological chain through selective access to light-matter interaction with specific physical sites. In particular, we present a comprehensive Density Matrix Renormalization Group (DMRG) study of ground-state features in different chain-cavity coupling geometries and validate analytical insights in the strong coupling regime. This new type of Majorana fermion correlation generation process comes with emergent cavity photon features. Moreover, by considering the time evolution after a sudden quench of the cavity-matter coupling strength, we show that the development of high non-trivial matter (Majorana) correlations leaves behind measurable non-classical photon imprints in the cavity. Innovative ways to dynamically generate TQ nonlocal correlations in topological chains of arbitrary length are thus provided, opening alternative routes to controllable long-range entanglement in hybrid photonic solid-state systems.

\end{abstract}
\maketitle

\section{Introduction}
The quantum matter realm, full of intriguing properties and potential applications, relies heavily upon the generation and control of highly nonclassical target states of matter. Recent years have seen tremendous progress in manipulating both natural~\cite{ yin2021probing} as well as synthetic~\cite{kim2018toward_MajoranaMagneticAtomicchains, ozawa2019topological,kiczynski2022engineering} topological systems. The possibility of managing nonlocal/extended states with a high level of  immunity to decoherence effects in topological systems is a key feature of paramount importance. A first step towards that goal is to reach the full coherent operation of nonlocal qubits from which topological quantum information/computing should follow. Either the Kitaev chain~\cite{kitaev}/ transverse Ising model~\cite{gomez_prb_18, huang_pan,Lukin_IsingSystems} provide adequate theoretical paradigms for supporting isolated Majorana fermions (MF) at their edges in the topological phase. Therefore, we will refer to both by Majorana chain (MC). The Kitaev model is well adapted to describe a strong spin-orbit semiconducting nanowire in proximity to a p-wave superconducting material where an axial applied magnetic field can tune the system into a topological phase. However, current superconducting qubit technology is already a mature platform for experimentally obtaining and manipulating Majorana fermions \cite{huang_pan,Majoranas_47_Qubits_science,ranvcic2022exactly_SimulationOnQuantumComputer}.  Circuit QED open boundary Ising-like realizations have amply been considered \cite{viehmann_13,zhang2014quantum_ExperimentIsingCQED,kjaergaard2020superconducting_SuperconductingCircuitsISING} and more recently, the localization properties of the MZEMs and their robustness against noise have been experimentally addressed~\cite{Majoranas_47_Qubits_science}. Braiding of non-abelian anyons \cite{Braiding_VortexMaj,Braiding_Cavity_Trif,Majorana_Magic} in such systems would be the grail of a large body of present day active research in condensed matter/quantum information physics. 

A promising path for such a purpose is provided by exploiting the interplay between cavity and matter systems. Recently, results on braiding and quantum teleportation of MFs have been shown in MC systems controlled by superconducting cavities \cite{huang_pan}. The field of cavity quantum materials with global light-matter coupled systems has shown that in the so-called light-matter deep-strong coupling (DSC) regime, new emerging properties appear in a wide range of condensed matter systems. The rich phenomena include the emergence of superconducting states~\cite{schlawin2022cavity}, manipulation of topological matter through cavity-matter interactions \cite{DmytrukSSH}, and many other potential applications \cite{kockum2019ultrastrong}. Regarding cavity interactions with topological matter, it is possible to manipulate systems featuring the so-called Majorana zero-energy modes (MZEM) embedded in a QED cavity~\cite{cai, mendez_prr, ricco2022accessing, platero, contamin}. Previously, a MC completely and uniformly placed inside a cavity in the weak coupling regime has been addressed~\cite{mendez_prr}. It was found that the R\'enyi entropy of the light-matter system can be directly connected with photon observables, a direct consequence of the fact that the quantum state of the cavity can be associated with a gaussian description. On the other hand, single qubit-resonator systems have been predicted to provide fast and high-fidelity qubit readout \cite{GrismoPRB}. Moreover, they are already used in superconducting quantum processors \cite{huang_pan}. Furthermore, the embedding of single qubits in a cavity has been widely studied \cite{PhysRevB.84.184515_qubitCavityCoherentControl, huang_pan,PhysRevA.105.062436_TriqubitnCavities} and shown to be useful for accessing the properties of MZEM~\cite{ricco2022accessing}. Thus, this plethora of exciting results motivate the present research.

The main goal of this work is to assess the ability of selective cavity QED set ups to create MZEMs out of the edges of a MC. Specifically, we show that, through strong light-matter interactions, it is possible to either shift non-local Majorana qubits from the edges of a topological chain or create additional pairs of non-local MF qubits by selectively embedding sections of the chain within a cavity. New features arising from MC-cavity strongly coupled systems do not only manifest in statics but also in dynamical aspects. We investigate both the ground-state (GS) properties as well as the time evolved states under a post-quench event with parameters covering a wide region of chemical potentials and coupling strengths, corroborating that emerging new bulk MF correlations turn out to be susceptible of manipulation by selectively coupling to a cavity. We show that such cavity induced MF correlation generation lies in an abrupt transition from an empty state to a regime where the MC-cavity GS shares both matter excitations and a cavity coherent photon state.  This transition leaves a characteristic light-matter bipartite entanglement that may be taken as one of its fingerprints. On the other hand, post-quench dynamical evolution shows transient non-classicality as measured by the photon Fano factor and the development of quantum state orthogonality as quantified by the Loschmidt echo. The site selective MC-cavity coupling procedure proves to be a remarkable tool for manipulating topological matter states. We analytically predict the value of the coupling strength for which the GS of the system switch between those states a result that is further endorsed by DMRG numerical calculations~\cite{SCHOLLWOCK,Or_s_2019DMRG}.

\begin{figure}[t!]
\begin{center}
 \includegraphics[width=1.0\linewidth]{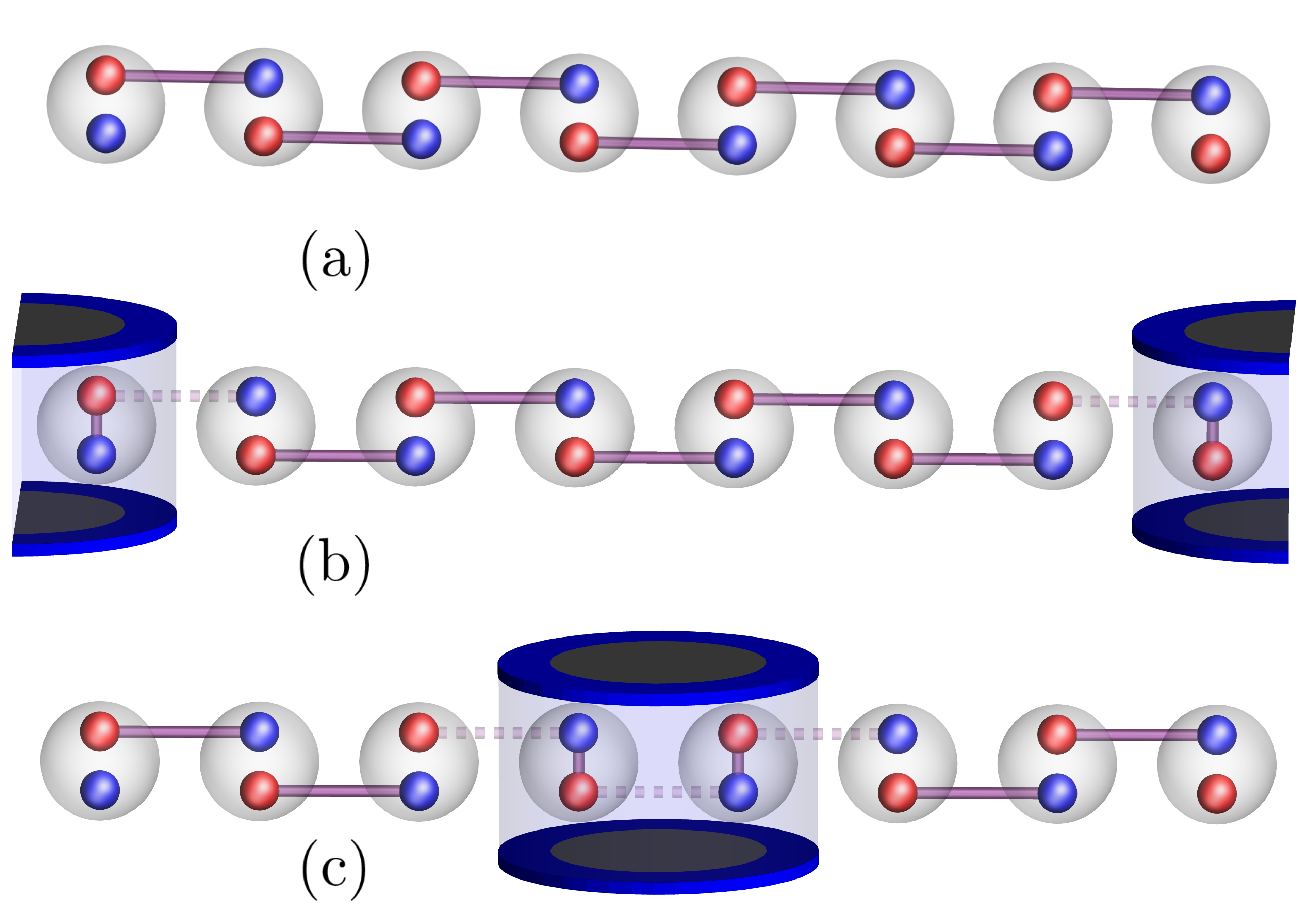}
\end{center}
  \caption{ Sketch of MC-cavity configurations at $\mu=0$. The shaded gray spheres represent MC real lattice sites to which a pair of MFs are allocated, depicted by blue ($\gamma_{j,1}$) and red small spheres ($\gamma_{j,2}$). Solid bonds portray the connectivity of MFs. Dashed lines represent bonds deteriorated by cavity-chain interaction. (a) The MFs at neighboring sites form $d$-fermions. In contrast, unbounded Majoranas at the ends encode free $d$-fermions, the so-called zero-energy modes.  (b) Edge geometry: the two ends of the open MC are embedded in the {\it same} cavity.  (c) Bulk geometry: two bulk sites of the open MC are embedded in the cavity. In all cases displayed, the cavity's effect is the fading of nonlocal $d$-fermions in favor of establishing local $c$-fermions.}
  \label{fig:Sketch}
\end{figure}

\section{General framework}\label{Sec1}
We consider a general radiation-matter framework consisting of a MC interacting with a single quantum radiation mode within a QED cavity. The Hamiltonian of the system reads as: 
$\hat{H}=\hat{H}_{\rm MC}+\hat{H}_{\rm C}+\hat{H}_{\rm X}$.
The $L$-site open boundary MC system Hamiltonian can be written in terms of spinless fermions as ($\hbar=1$)
\begin{equation}
\begin{split}
\hat{H}_{\rm MC}=&-\frac{\mu}{2} \sum_{j=1}^{L}\left[2\hat{c}_{j}^\dagger \hat{c}_{j}-\hat{1}\right]\\
           &-\Delta\sum_{j=1}^{L-1}\left[\hat{c}_{j}^{\dagger}\hat{c}_{j+1}+\hat{c}_{j+1}^{\dagger}\hat{c}_{j}-\hat{c}_{j}\hat{c}_{j+1}-\hat{c}_{j+1}^{\dagger}\hat{c}_{j}^{\dagger}\right],
\end{split}
\end{equation}
where $\hat{c}_{j}$ (${\hat{c}_{j}^{\dagger}}$) is the annihilation (creation) operator of spinless fermions at site $j=1,\ldots, L$ with anti-commutator $\{\hat{c}_{j}, \hat{c}_{j^{\prime}}^{\dag} \}=\delta_{j,j^{\prime}}$, $\mu$ denotes the chemical potential, $\Delta$ the hopping amplitude between nearest-neighbor sites and the nearest-neighbor two-fermion pairing interaction taken identical to $\Delta$ for simplicity (we assume $\Delta\geq 0$ without loss of generality). Within this homogeneous regime (which we will keep throughout the paper), the MC features two phases: a topological and a trivial phase. In the former phase a MZEM emerge whenever $|\mu| < 2\Delta$~\cite{majorana, kitaev}.

A remarkable fact of this 1D lattice model is the possibility of expressing each physically ({\it localized}) real lattice fermion (we call them as local $c$-fermions) in terms of a pair of MFs, $\gamma_{j,1}$ and $\gamma_{j,2}$. Moreover, depending on the pairing scheme chosen for these MFs, one can build new {\it delocalized} fermions~\cite{leumer2020exact}. In particular, in the deep topological phase at $\mu=0$, the MC Hamiltonian becomes
\begin{equation}
\hat{H}_{\rm MC}=
i\Delta\sum_{j=1}^{L-1}\gamma_{j,2}\gamma_{j+1,1}
=2\Delta\sum_{j=1}^{L-1}\left[\hat{d}_{j}^{\dagger}\hat{d}_{j}-\frac{1}{2}\right]
\label{Eq:n1}
\end{equation}
where $\gamma_{j,1}=\hat{c}_{j}^{\dagger}+\hat{c}_{j}=i(\hat{d}_{j-1}^{\dagger}-\hat{d}_{j-1})$ and $\gamma_{j,2}=i(\hat{c}_{j}^{\dagger}-\hat{c}_{j})=\hat{d}_{j}^{\dagger}+\hat{d}_{j}$ with $j=1,..,L-1$. Two important consequences follow from Eq.~\eqref{Eq:n1}: (i) the energy eigenstates of this MC are composed of integer number of fermionic quasi-particles denoted by $\hat{d}_{j}^{\dagger}$, ($\hat{d}_{j}$) instead of real fermions (we call them as non-local $d$-fermions). (ii) The Majorana fermion $\gamma_{1,1}$ at the left end and the Majorana fermion $\gamma_{L,2}$ at the right end are missing in the diagonalized Hamiltonian. Physically, this means that there are two {\it isolated} MFs, $\gamma_{1,1}$ and $\gamma_{L,2}$, which are localized at opposite ends with zero eigenenergy, the so-called MZEM. From these uncoupled MFs it is still possible to form a highly non-local $d$-fermion as $\hat{d}_{0}=\frac{1}{2}\left ( i\gamma_{1,1}+\gamma_{L,2} \right )$. The above discussion is schematically illustrated in Fig.\ref{fig:Sketch} (a).

The single mode microcavity radiation Hamiltonian takes the usual form
$\hat{H}_{\rm C}=\omega \hat{a}^\dagger\hat{a}$,
where $\omega$ is the photon quantum energy, $\hat{a}^\dagger$ and $\hat{a}$ are the creation and annihilation photon operators, respectively. Finally, $\hat{H}_{\rm X}$, corresponds to the MC-microcavity coupling terms which reads as,
$\hat{H}_{\rm X}=\frac{\lambda}{\sqrt{n_{\rm Cav}}}\left ( \hat{a}^\dagger+\hat{a} \right )\sum_{j\in {\rm Cav}} \hat{c}_{j}^\dagger \hat{c}_{j}$,
where $\lambda$ is the coupling strength between matter and radiation subsystems (without loss of generality $\lambda>0$), ${\rm Cav}$ is the set with the site indices interacting with the cavity, and $n_{\rm Cav}:=|{\rm Cav}|$. Linear  coupling schemes have been employed extensively in various studies~\cite{MirceaTrifHamiltonian,TrifMajoranasSpinOrbit,mendez_prr,CottetPRX_Coupling,Cottet_Vacumm_Ham,Trif_QuantumDot,UltraStrong_superconducting,Dmytuk_GaugeFixing}. However, all of the previous investigations were concerned with a homogeneous regime where the cavity couples uniformly with all sites in the MC. In the following, we examine the effects of a selective cavity which directly couples only to $n_{\rm Cav} < L$ sites of a chain partly trapped in it. On the experimental side, the physical realization of superconducting microcavities affecting only selected elements of a MC has been amply discussed in Refs.~\cite{viehmann_13,huang,ricco2022accessing}.

One of the main results we address extensively below is that the effect of the cavity on the MC is the transformation of nonlocal $d$-fermions into local $c$-fermions (horizontal bonds vs. vertical bonds in Fig. \ref{fig:Sketch}) redrawing in this way the MF connectivity. We confirm this fact by computing both MC and photon linked features in strongly coupled MC-cavity systems, finding that the likelihood of freeing up bulk MFs along a MC is allowed. If the edges of the open chain interact with the same local cavity (a kind of folded chain in the form of a ring), we will call the resulting configuration as {\it edge geometry}, while the term {\it bulk geometry} will indicate a cavity embedding only bulk chain sites (see Fig. \ref{fig:Sketch}). On one hand, for the edge geometry, the annihilation of the nonlocal $d$-fermion qubit in favor of a local $c$-fermion qubit inside the cavity creates free MF in the adjacent neighbor sites outside the cavity (chain bulk). On the other hand, for the bulk geometry case, an additional pair of free MF does appear just outside the cavity, corresponding to effectively cutting the original chain in two shorter MC. This annihilation/creation process of long ranged bulk nonlocal qubits is a required feature to perform braiding operations in T junctions \cite{Pachos_Lahtinen_2017_Braiding}. 

MF correlations are captured by the two-point correlator $Q(i,j)$, defined as $Q(i,j)\equiv 2|\langle \hat{c}_i \hat{c}_j^\dagger+\hat{c}_j \hat{c}_i^\dagger \rangle|=|\langle \gamma_{i,2}\gamma_{j,1}-\gamma_{i,1}\gamma_{j,2} \rangle|$. 
The two-point correlators, also known as string order parameters, take values 
different from $0$ in the topological phase \cite{Vishwanath_two-point-correlatios}. Furthermore, these correlations are maximal when evaluated at the points corresponding to the defects that permit the emergence of a topological order \cite{Pollman_twoends}. For an infinitely long isolated MC, when calculated at the edges, $Q(1,L)=1$ in the topological phase while $Q(1,L)=0$ in the trivial phase \cite{ReslenQ}. However, for finite size chains, $Q$ lies between those values, being close to $1$ when the MZEM are completely localized at the boundaries. In particular, at $\mu=0$, an isolated MC [see Fig. \ref{fig:Sketch}(a)] should reach maximum two-point correlation for sites $i$
and $j$ corresponding to the same $d$-fermion qubit and $0$ otherwise, i.e. $Q(i,j)=\delta_{|i-j|,1}$ for non-zero energy modes, and $Q(1,L)=1$ for the MZEM $\hat{d}_{0}$-fermion. Consequently, the two-point correlators show as excellent indicators of non-locality, closely connected to the main topological features of the system \cite{LeeQ}. Thus, the main focus will be put on $Q(i,j)$, with indices $i$ and $j$ that are going to be varied across the MC to determine the eventual creation of bulk free MF or equivalently new MZEMs.

\begin{figure}[t!]
\begin{center}
 \includegraphics[width=1.0\linewidth]{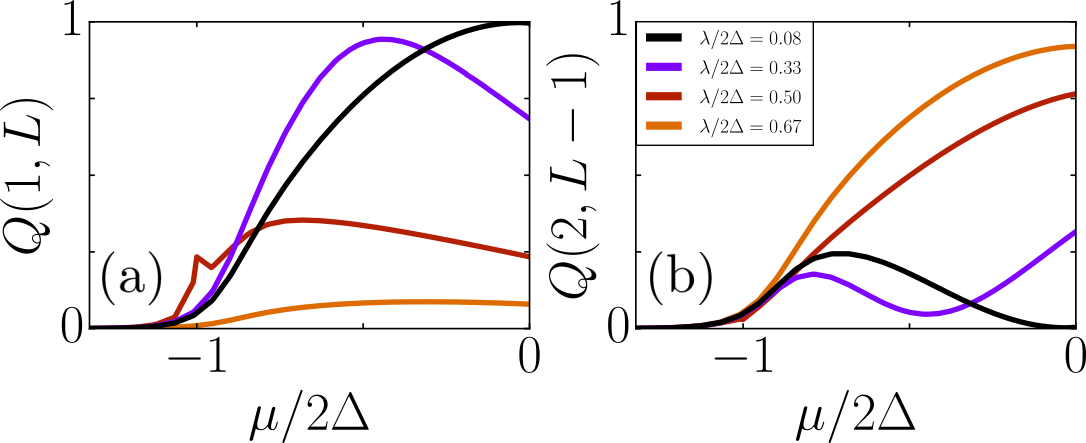}
\end{center}
\caption{Edge geometry. $Q(i,j)$ correlations showing the displacement of the weights from the edges to the bulk as the coupling increases.  (a) MC edge-edge $Q(1,L)$ correlation, and (b) correlation for nearest neighbors sites of the edges, $Q(2,L-1)$. The parameters of the simulated system are $L=20$, $n_{\rm Cav}=2$ and $\omega/2\Delta=0.25$.}
  \label{fig:Q1L}
\end{figure}

\section{Ground-state results}
We unveil the localization shifting signatures of nonlocal MF-qubits in the whole GS for selective coupling of the MC to a cavity, complementing paradigmatic global coupling results.

\begin{figure}[t!]
\begin{center}
 \includegraphics[width=1.0\linewidth]{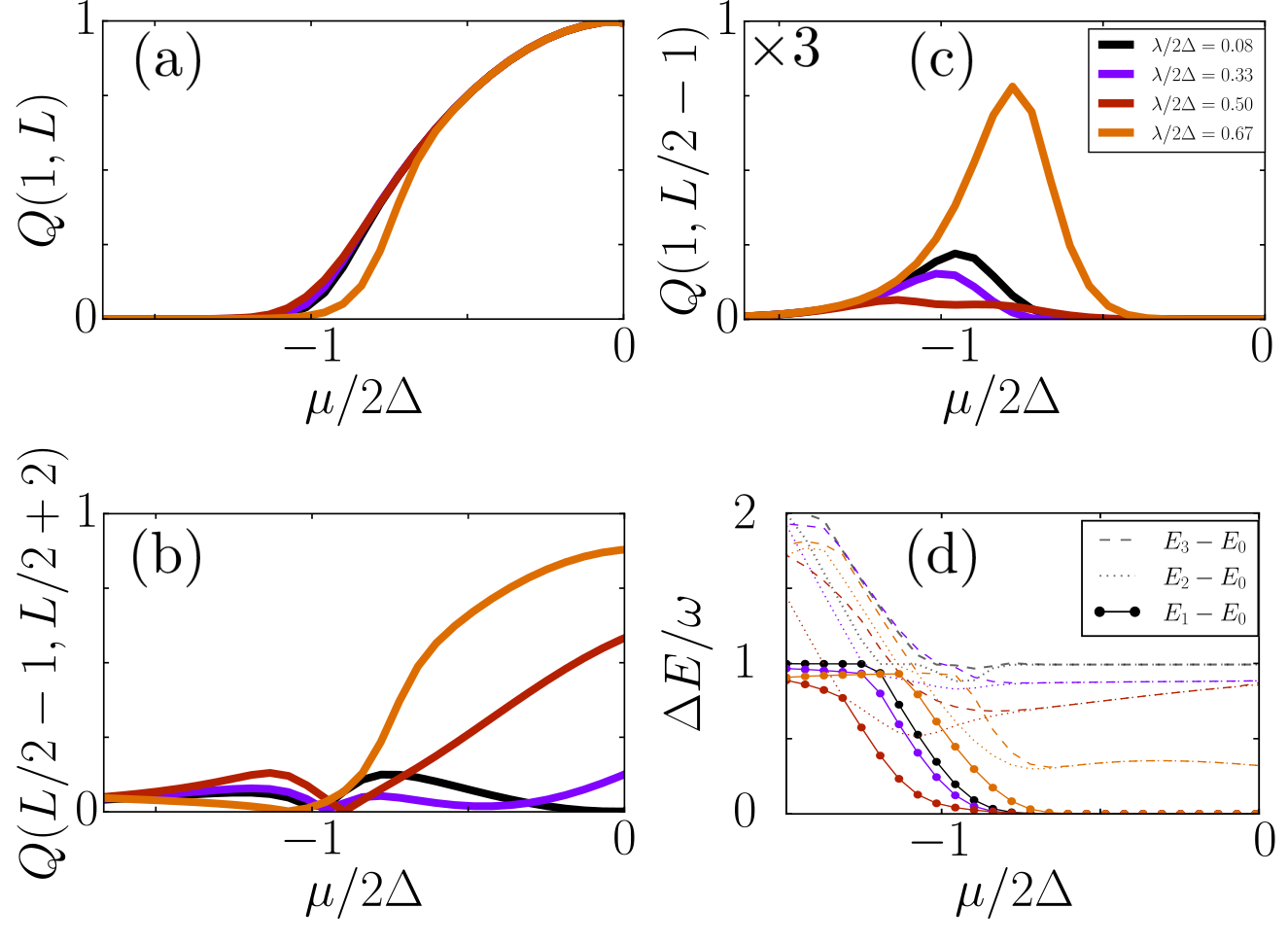}
\end{center}
\caption{Bulk geometry.  $Q(i,j)$ correlations as a function of the chemical potential displaying the emergence of an additional pair of dominant correlations at bulk locations as the coupling increases. (a) MC edge-edge  $Q(1,L)$ correlation. (b)  $Q(L/2-1,L/2+2)$ correlation between the external first-neighbors to the sites inserted into the cavity. (c) $Q(1,L/2-1)$ correlation multiplied by three. (d) Energy differences between the first (dashed line), second (solid line), and third (dotted line) excited states with respect to the ground-state. Same parameters as in Fig. \ref{fig:Q1L}. }
  \label{fig:Bulk_all}
\end{figure}

Let us start with the limit $\mu\rightarrow 0$ together with the mean-field picture framework (see Appendix \ref{Sec:MF}). For an isolated MC in this limit, MF are paired to form nonlocal $d$-fermions [see Fig.\ref{fig:Sketch} (a)]. However, as already mentioned, the main effect of the cavity is to transform those nonlocal fermion modes into local $c$-fermions involving sites $j \in {\rm Cav}$. This process has as its main consequence the decoupling of MF that were initially bounded non-locally to sites embedded in the cavity.
Accordingly, at MC-cavity couplings large enough such that this transformation is complete, we expect new free MF to appear at the neighboring MC sites outside the cavity. This behavior is confirmed by depicting $Q$-correlations for both edge [see Figs.~\ref{fig:Q1L}(a)-(b)] and bulk geometries [see Figs.~\ref{fig:Bulk_all}(a)-(c)], as a function of the chain chemical potential $\mu$. At $\mu\rightarrow 0$, for edge geometry, $Q(1,L)\approx 1$ while $Q(2,L-1)\approx 0$ for weak enough MC-cavity coupling strength. However, this behavior can be completely reversed at strong coupling, i.e., $Q(1,L)\rightarrow 0$ while $Q(2,L-1)\rightarrow 1$, [see Fig. \ref{fig:Sketch}(b)]. On the other hand, for bulk geometry, $Q(1,L)\approx 1$ for any coupling strength. The latter emphasizes the robustness of the zero energy $d$-fermion mode, or qubit, formed by MFs at the chain edges irrespective of the cavity presence. Interestingly, the correlation between cavity-generated new bulk MZEM, $Q(L/2-1,L/2+2)$, grows from $0$ for weak coupling to near $1$ at strong coupling, [see Fig. \ref{fig:Sketch}(c)].

Although a more complex behavior is expected at finite chemical potentials, $\mu\neq 0$, the essential physics remains unchanged. The only difference is that the MF quantum state broadens to spread over several lattice sites, instead of the ideal pairing localization found at $\mu=0$. In the following, we only present observable features for $\mu<0$ since, in this region, significant differences between the isolated and cavity-coupled MC cases are more evident. However, results for $\mu\geq0$ can be found in the Appendix \ref{sec:Robustness} and they complete the spectrum for experimentally accessible $\mu$ parameters (cf. Refs. \cite{huang_pan, Majoranas_47_Qubits_science}). All the static results presented below were obtained by DMRG simulations.

For edge geometry, Fig.~\ref{fig:Q1L}(a), $Q(1,L)$ as a function of $\mu$ shows a maximum shifting towards increasingly negative chemical potential values. At a certain strong coupling strength a spike is visible, an indicator of the transition between different GS as will be discussed below. The results shown in Figs. \ref{fig:Q1L}~(a)-(b) demonstrate a rapid decreasing of long range MC correlations as $\mu/2\Delta \rightarrow -1$, consistent with approaching the border of the MC phase diagram indicating the transition from the topological phase to the trivial phase. As the coupling strength $\lambda$ increases, the weights of the MZEM in the correlation functions are shifted from the edges to their nearest neighbor sites, representing the effective downgrading of the nonlocal $d$-fermion inside the cavity, which becomes transformed into local $c$-fermions. Consequently, new free MFs emerge at the neighboring sites of the cavity, confirming that it is possible to move around MF based non-local qubits, an effect that could be advantageous for implementing cavity enhanced braiding protocols.  

Fig. \ref{fig:Bulk_all} shows bulk geometry correlations for $\mu\le 0$ and different pairs of sites $i$ and $j$. 
In the MC topological region, close to the phase transition at $\mu/2\Delta=-1$, $Q(1,L/2-1)$ becomes the dominating correlation in the strong MC-cavity regime  [$Q(L/2+2,L)=Q(1,L/2-1)$ due to spatial symmetry]. Therefore, two pairs of indices define two $d$-fermions with a bulk MZEM. In contrast, for smaller values of $|\mu|$, far from the phase transition point, the dominant correlation is given by $Q(1,L)$ as should be for the isolated original MC regardless the presence of a bulk inserted cavity. However, new correlations are arising. Specifically, $Q(L/2-1,L/2+2)$ increases in consistency with the generation of new free MF in the bulk sites just outside the cavity. This behavior can be interpreted as effectively separating the sections interacting with the cavity from the rest of the system. When strongly coupling the electromagnetic mode to a segment of bulk sites, the original chain effectively transforms into two weakly interacting MCs. Each MC will provide a pair of free MFs at their edges [effective ``edges in the bulk" as it is also supported by a mean-field description (see Appendix \ref{sec:QforBulk})].

Evidence for the establishment of a new MZEM in the bulk geometry is presented in Fig.~\ref{fig:Bulk_all}(d). For low coupling strengths, the GS degenerates in the topological phase. In addition, the mode is gapped from the second and third energy levels. As the MC-cavity coupling is increased, the first and second excited states become degenerated with the GS, representing the progressive creation of additional free MFs.

\begin{figure}[t!]
\begin{center}
 \includegraphics[width=1.0\linewidth]{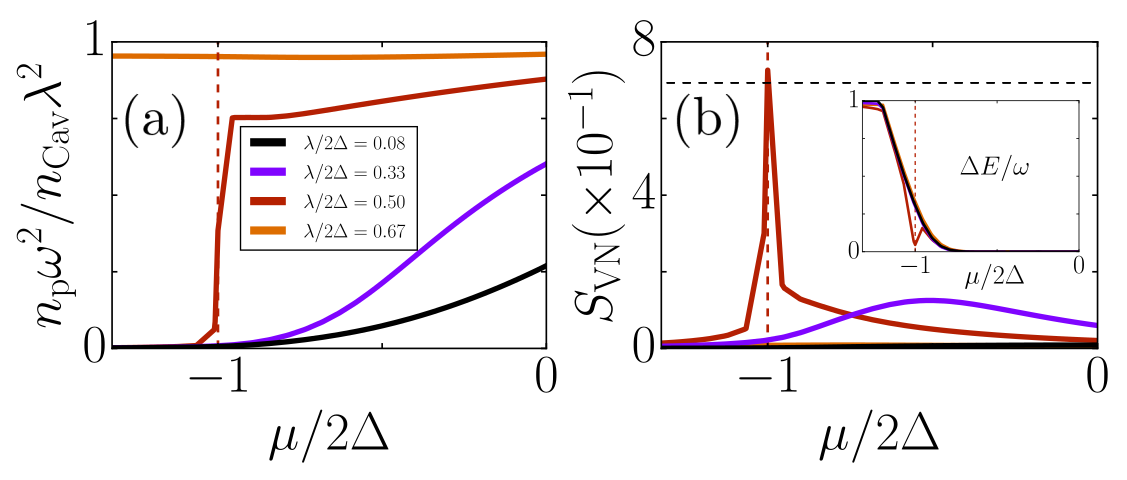}
\end{center}
\caption{Edge geometry. (a) Number of photons ($n_{\rm p}$) in the cavity. (b) MC-cavity Von Neumann entropy ($S_{VN}$). The dashed black line represents the value of $S_{VN}=\ln(2)$ which is the entanglement of two states with equal probability. (Inset) Energy gap from the ground-state to the first excited state of the light-matter interacting system. The dashed red line represents the predicted point of the transition in the strong-coupling limit (see main text). Same parameters as in Fig. \ref{fig:Q1L}. }
  \label{fig:VN1}
\end{figure}

Now, we address the main photon features and how they correlate with $d$-fermion unpairing processes. First, the cavity mean photon number is displayed in Fig. \ref{fig:VN1}(a) providing evidence of the interplay between the cavity photon occupation and the generation of new free MFs. In order to realize that the cavity privileges the formation of $c$-fermions, it is fruitful to distinguish the states that compete to dominate the GS physics. The nature of the GSs becomes clear through Fig. \ref{fig:VN1}(a). For a negative enough chemical potential, the GS becomes separable corresponding to a product between the GS of the isolated MC and a vacuum state of the radiation field, $|0_{MC} \rangle \otimes |0_{ph} \rangle$, as it will be shown below. In contrast, for $\mu\rightarrow 0$, a large number of photons can occupy the cavity. This distinction is more abrupt as the value of the light-matter coupling strength gets large enough.

The aforementioned change in the nature of the GS can be simply understood in the DSC limit. By considering $\hat{H_0}=\hat{H}_{\rm X}+\hat{H}_{\rm C}$ as the leading term in the Hamiltonian, the eigenstates are given by

\begin{equation}
    \ket{\phi_{\nu,n}}=\hat{D}\pap{-\frac{\lambda\nu}{\sqrt{n_{\rm Cav}}\omega}}\ket{\nu}\otimes\ket{n}
\end{equation}
with corresponding eigenenergies given by $ E_{\nu,n}=\omega n - \frac{\nu^2\lambda^2}{n_{\rm Cav}\omega}$. In the latter expressions $\hat{D}(\alpha)=\exp\pas{\alpha \hat{a}^\dagger - \alpha^* \hat{a}}$ is the displacement photon operator and $\ket{\nu}$ is an element of the basis set that diagonalizes the on-site fermion occupation term $ \sum_{j\in {\rm Cav}} \hat{c}_{j}^\dagger \hat{c}_{j}$ with $\nu \in \{0,1,...,n_{\rm Cav}\}$ the corresponding eigenvalue.
Obviously, the GS energy will be found when $n=0$ and a maximum filling of MC $c$-fermions embedded into the cavity ($\nu=n_{\rm Cav}$) is reached. Remarkably, the expected number of photons $\langle{\phi_{\nu,n}}|\hat{a}^\dagger\hat{a}|{\phi_{\nu,n}}\rangle=n+\frac{\nu^2\lambda^2}{n_{\rm Cav} \omega^2}$ is proportional to $\nu^2$. This interplay between the cavity and $c$-fermion occupation defines the main physical features of the coupled MC-cavity system. This mutual behavior implies that when the cavity is maximally displaced, the sites $\in {\rm Cav}$ act as filled $c$-fermions. Next, we consider the isolated MC term $\hat{H}_{\rm MC}$ as a perturbation. In the limit $\mu/2\Delta \ll -1$, and in first order in perturbation theory, $\hat{H}_{\rm MC}$ will make corrections to the unperturbed GS energy
of the form (see Appendix \ref{sec:firstOrderPerturbation})
\begin{equation}
    E_{\rm DSC}=E_{\rm MC}-n_{\rm Cav}\mu-\frac{n_{\rm Cav}\lambda^2}{\omega}
\end{equation}
with $E_{\rm MC}$ the GS energy of the isolated MC. Therefore, the GS will be associated with a cavity photon displacement and filled of $c$-fermions when $E_{\rm DSC}<E_{\rm MC}$. On the contrary, the GS corresponds to a separable state product of no fermion excitations in the MC and a vacuum state for the cavity. Thus, the transition between both sectors of GSs occurs at $\mu_c=-\lambda^2/\omega$ which matches fairly well with the spike in $Q(1,L)$ for $\lambda/2\Delta=0.5$ in Fig.~\ref{fig:Q1L}(a), the onset of photon excitations and the von Neumann entropy peak in Figs. \ref{fig:VN1}(a)-(b), respectively. Hence, we propose these findings for detecting the transformation from nonlocal $d$-fermion to local $c$-fermion through cavity observables. The latter adds to the methods for characterizing topological phases with cavity measurements \cite{mendez_prr,Exp_Cav_Werner}.

Further support of our description of the transition is provided by the inset in Fig.~\ref{fig:VN1}(b), which shows that an anti-crossing of energy levels occur at the transition points when $E_{\rm MC}$ and $E_{\rm DSC}$ approach the same value. This anti-crossing endorses us with a transition signature, namely the Von Neumann entropy ($S_{\rm VN}$), which is shown in Fig.\eqref{fig:VN1}(b). At the transition, $S_{\rm VN}$ is near $\ln(2)$, which is a typical value for a diagonal reduced density matrix for a mixed cavity state in which only two pure states with equal probability have the dominant weights. From previous explanations above, it is clear that the reduced density matrix is approximately describing a pair of orthogonal cavity states with $0$ and $(\nu \lambda)^2/(n_{\rm Cav} \omega^2)$ photons. Additionally, note that the GS in the topological phase still degenerates regardless of the interaction with the cavity. The degenerancy proves the ability of the light-matter GS to encode qubit states from the occupation of the $d$-fermions sitting at the new effective MC edges. Similar features appear for the bulk geometry regarding the $S_{\rm VN}$ (see Appendix \ref{sec:Robustness}), and anti-crossing. Then this system is another example where light-matter entanglement signals GS transitions~\cite{mendez_prr,Chanda}. However, the main difference between edge and bulk cavity physics is the creation of a new free $d$-fermion for the latter. We emphasize that, although the results displayed here are for a given $L$ and $n_{\rm Cav}=2$, the mechanism presented is robust to the chain length and the number of sites embedded in the cavity (see Appendix \ref{sec:Robustness}). Furthermore, our results are robust against inhomogeneities in the cavity coupling for different sites.

\section{Time-dependent results}
The local and non-local properties of MF based qubits just described, manifest themselves in the system's dynamics too. We proceed by focusing on the cavity induced dynamical aspects of the generation of isolated free bulk MFs by considering the experimentally relevant quenching process: at time $t=0$ a MC empty of fermion excitations GS is suddenly coupled to a cavity in its vacuum state, i.e. $|\Psi(0) \rangle = |0_{MC} \rangle \otimes |0_{ph} \rangle$. Here, we limit ourselves to the analysis of the simpler sudden switching on case in order to extract valuable physical information on the ultra-fast growing up of new cavity induced fermion correlations and their main photon twin markers which could be experimentally accessible in current realizations of MC-cavity systems.

At $\mu=0$, only four MC states are affected by the cavity: $\{  |0_{MC} \rangle,$  $\hat{d}_0^{\dagger}\hat{d}_1^{\dagger}|0_{MC} \rangle,$ $ \hat{d}_{L-1}^{\dagger}\hat{d}_0^{\dagger}|0_{MC} \rangle,$ $ \hat{d}_{L-1}^{\dagger}\hat{d}_1^{\dagger}|0_{MC} \rangle \}$ for edge geometry, while for a bulk geometry with ${\rm Cav}=\{j, j+1\}$  the states are $\{  |0_{MC} \rangle,$ $  \hat{d}_j^{\dagger}\hat{d}_{j+1}^{\dagger}|0_{MC} \rangle,$ $ \hat{d}_{j-1}^{\dagger}\hat{d}_{j+1}^{\dagger}|0_{MC} \rangle,$ $ \hat{d}_{j-1}^{\dagger}\hat{d}_j^{\dagger}|0_{MC} \rangle \}$. Thus, time dependent results are the same regardless the MC length $L$ for each considered MC-cavity geometry.

\begin{figure}[tbh!]
\includegraphics[width=1.0\linewidth]{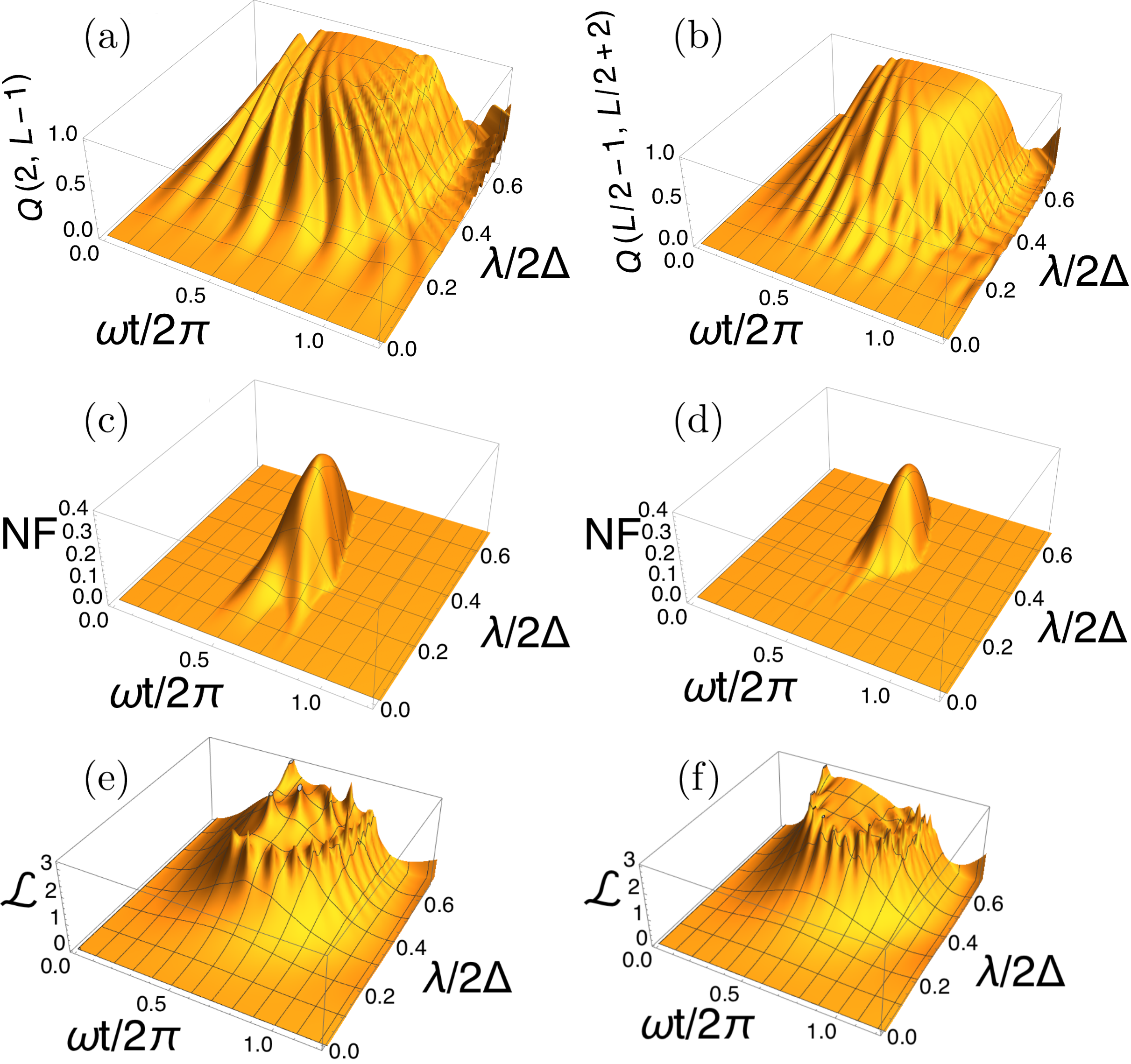}
\caption{ Left column: edge geometry. Right column: bulk geometry. (a)-(b) Emerging bulk MF correlations $Q(2,L-1)$ and $Q(L/2-1,L/2+2)$, (c)-(d) modified NF-Fano factor (see main text) and (e)-(f) Loschmidt echo. All plots are function of the dimensionless time ($\omega t/2\pi$) and have light-matter coupling strength ($\lambda/2\Delta$). The cavity frequency is $\omega/2\Delta=0.25$ and $\mu=0$.}
 \label{fig:time}
\end{figure}

The time evolution of the relevant cavity-induced emerging correlations previously discussed for edge and bulk geometries as a function of time and MC-cavity coupling strength $\lambda$ are shown in Figs. \ref{fig:time}(a)-(b), respectively. As expected, at $t=0$ both correlations vanish as the considered sites do not hold free MFs. As time progresses the correlations increase reaching their maximum values near the half a cavity period, i.e. at $\omega t/2\pi \approx 1/2$. This correlation building up effect is significant for strong coupled MC-cavity systems.

Moreover, it is interesting to see the dynamical connection between MC-correlations with non-classical photon features through the Fano factor, ${\rm FF} = \pap{\langle \hat{n}^{2} \rangle - \langle \hat{n}\rangle^{2}}/\langle \hat{n}\rangle$ with  (${\rm FF}<1$ for a non-classical or sub-Poissonian photon state \cite{moreau2019imaging}). In order to highlight the photon state non-classicality, Figs. \ref{fig:time}(c)-(d) display the quantity ${\rm NF}=(1-{\rm FF})\Theta(1-{\rm FF})$ as a function of time and MC-cavity coupling strength [$\Theta(x)$ is the Heaviside function]. It is evident that reaching a maximum value for non-trivial bulk matter correlations $Q(2,L-1)$ and $Q(L/2-1,L/2+2)$ occurs simultaneously with the emergence of non-classical photon states in the cavity for both considered MC-cavity geometries. NF non-classical features not only develop earlier for lower coupling strengths but they also get higher for the edge geometry case, indicating a stronger coupling of edge Majorana fermions with the selective cavity photon mode. This sub-Poissonian feature in the photon state is a pure transient effect that does not appear in the GS properties of the whole system previously discussed. Thus, light-matter correlated transient behaviors could be exploited for designing protocols aiming to optically generate and detect bulk MZEM in a MC-cavity system.

On the other hand, and intimately linked with the FF behavior just described, the Loschmidt echo (LE), quantifying the overlap of an evolved state with the initial state, defined by ${\cal L}(t)=-log_{10}|\langle \Psi(0)|\hat{U}(t,0)|\Psi(0) \rangle|$, where $\hat{U}(t,0)$ is the time evolution operator, is shown in Figs. \ref{fig:time}(e)-(f). For strong enough MC-cavity coupling  ${\cal L}(t)$ reaches a high plateau, coming with sparse peaks, indicating the whole system attains a quasi-orthogonality condition with respect to the initial state for  both considered MC-cavity geometries.  Once more, the development of high matter non-trivial correlations coincides with the presence of orthogonality peaks in the LE~\cite{Loschmidt_Betancourt}. 

These time-dependent results should still remain valid for MC chemical potentials $\mu\approx 0$ where the hopping to different $d$-fermions non directly connected with the cavity would be feasible. 
\section{Conclusions}
In summary, we have found that strongly coupling QED-light with quantum topological matter at select sites and different geometries, can efficiently generate MF-based delocalized qubits. By selectively controlling the spatial distribution of the light-matter interaction, it is possible to create setups where the connectivity of MF can be tailored. The latter possibility arises since the on-site light-matter interaction restricts the filling of the coupled chain sites to be proportional to the photon occupation of the cavity mode. We have shown that for an edge geometry, it is possible to displace isolated MFs from the chain edges to inner bulk sites, and for the bulk geometry, new free nonlocal $d$-fermions in bulk sites emerge. In the DSC regime, the occupation of the sites coupled directly to the cavity are forced to saturate with one fermion per physical site as the interaction strength increases. Thus, the affected sites become effectively decoupled from the rest of the chain. 

Our dynamical findings provide a way to generate nonlocal correlations in MC of arbitrary length starting from a trivial empty state, allowing the creation of controllable long-range entanglement in hybrid photonic solid-state systems. Thus, these results offer QED-based new paths to explore avenues for future research on Majorana fermion braiding as well as on distinguishing the nonlocal Majorana pairing from the topologically trivial Andreev bound states.\\

In this manner, selective access to light-matter interaction with specific physical sites has proven to be a remarkable tool for manipulating topological matter. We highlight that the setup proposed here is highly relevant for applications in quantum superconducting circuits with 
modern technologies. More broadly, our results are an example of how spatially tailored light-matter interaction can produce a tight interdependence between light and matter systems.

\section*{ACKNOWLEDGMENTS}
The authors acknowledge F.J.G\'omez-Ruiz and J.J.Mendoza-Arenas for stimulating discussions at earlier stages of this research. DMRG calculations were perfomed with the TNT ~\cite{TNT,Sarah} and TeNPy~\cite{tenpy} libraries. F.P.M.M.-C thanks F.Schlawin for helpful comments. F.P.M.M.-C., F.J.R and L.Q are thankful for financial support from Facultad de Ciencias-UniAndes projects: INV-2021-128-2292, INV-2019-84-1841, INV-2022-137-2412 and INV-2022-144-2488, and Universidad de los Andes High Performance Computing (HPC) facility. C.T acknowledges financial support from the Agencia Estatal de Investigación of Spain, under contract PID2020-113445GB-I00, and Proyecto Sinergico CAM 2020 Y2020/TCS-6545 (NanoQuCo-CM).

\bibliography{Photon-Fermion}
\newpage
%
\appendix
\section{Mean-field approach}\label{Sec:MF}

The procedure presented here is similar to that in Ref. \cite{G&MMeanField}, and details have already been provided in Ref. \cite{mendez_prr}. The subject of the mean-field approximation will be the interaction part of the Hamiltonian $\hat{H}_{{\rm X}}$. For this, the quantum fluctuations of products of bosonic and fermionic operators are set to $0$, therefore
\begin{equation}
\pap{\hat{a}^\dagger+\hat{a}-\langle\hat{a}^\dagger+\hat{a}\rangle}\pap{\hat{c}_j^\dagger \hat{c}_j-\langle\hat{c}_j^\dagger \hat{c}_j\rangle}=0.
\end{equation}
Since we are only interested in the GS of the system, we can study the properties of the expected value of the mean-field Hamiltonian:
\begin{equation}
  \langle \hat{H}_{{\rm MF}}\rangle=\expv{\hat{H}_{\rm K}}+n_{\rm Cav} \omega x^2+2\lambda x C
\end{equation}
where expectation values are calculated with the photon-fermion GS. Here, we define:
\begin{equation}        x=\frac{\langle \hat{a}+\hat{a}^{\dagger}\rangle}{2\sqrt{n_{\rm Cav}}},\quad          C=\sum_{j \in {\rm Cav}}\expv{ \hat{c}_j^\dagger \hat{c}_j},
\end{equation}
the only difference with the derivation in Ref. \cite{mendez_prr} comes from the possibility of coupling only $n_{\rm Cav}$ sites instead of $L$. To obtain the mean-field Hamiltonian, one finds that the GS will be the product of the chain state with a coherent state $\ket{\sqrt{n_{\rm Cav}}x}$. Consequently, the photonic part of the Hamiltonian will describe a displaced harmonic oscillator, with photon number $\langle \hat{a}^\dagger \hat{a} \rangle\equiv \langle \hat{n} \rangle=n_{\rm Cav} x^2$.

The minimization of the expected GS energy with respect to $x$, $\partial \langle \hat{H}_{{\rm MF}}\rangle/ \partial x=0$, yields
\begin{equation}
\label{eq:Minimization}
 x=-\frac{\lambda}{\omega}\frac{C}{n_{\rm Cav}},
\end{equation}
which shows the interdependence of the cavity and chain observables. Therefore we can see that $x$ is restricted as $x\in [-\frac{\lambda}{\omega},0]$.

The mean-field Hamiltonian for the chain subspace provides an understanding of the configuration of dominating indices for  $Q(i,j)$. Let us assume that we know the value of $x$. Therefore, the effective Hamiltonian of the chain has the form
\begin{equation}
\label{eq:HChain_Effective}
    \hat{H}_{\rm MC}^{\rm MF}=\hat{H}_{\rm MC}+2\lambda x     \sum_{j \in {\rm Cav}} \hat{c}_j^\dagger \hat{c}_j.
\end{equation}
From this expression, it is clear that the coupling with the cavity modifies the chemical potential leading to the behavior described in the main text (MT) for DSC.

This mean-field Hamiltonian allows us to resort to the Bogoliubov solution of the MC model (details have already been presented in \cite{gomez_prb_18}). After the diagonalization, the Hamiltonian for a single isolated MC adopts the simple form
\begin{eqnarray}
\hat{H}_{\rm MC}^{\rm MF}=\sum_{k=1}^{L}\epsilon_k\left[\hat{d}_{k}^\dagger \hat{d}_{k}-\frac{1}{2}\right],
\label{Eq:n21}
\end{eqnarray}
where the new fermion operators, or quasi-particle operators, $\hat{d}_{k}$ and $\hat{d}_{k}^\dagger$ in Eq.~(\ref{Eq:n21}), obeying $\left \{\hat{d}_{k},\hat{d}_{k'}^{\dagger}\right \}=\delta_{k,k'}$, are connected with fermion site operators $\hat{c}_{j}$ through the Bogoliubov transformation $\hat{c}_{j} = \sum_{k=1}^{L} \left ( U_{k,j} \hat{d}_{k} + V_{k,j}\hat{d}^{\dagger}_{k} \right )$.
For a $L$ site MC, the coefficients $U_{k,j}$ and $V_{k,j}$ are determined by the eigenvectors of the Bogoliubov-De Gennes solution, which can be represented by a $2L \times 2L$ matrix \cite{gomez_prb_18}. The MC displays a spectrum with positive and negative energies, $\pm\epsilon_k$ with $k=1,...,L$. With the eigenvectors, $Q(i,j)$ reads simply
\begin{equation}
    Q(i,j)=4|\sum_{k=1}^LV_{k,i}V_{k,j}|,
\end{equation}
where $k$ represents the mode index and $i$, $j$ are spatial site indices. 

\section{Robustness of the description}
\label{sec:Robustness}
\begin{figure*}[t!]
\begin{center}
 \includegraphics[width=1.0\textwidth]{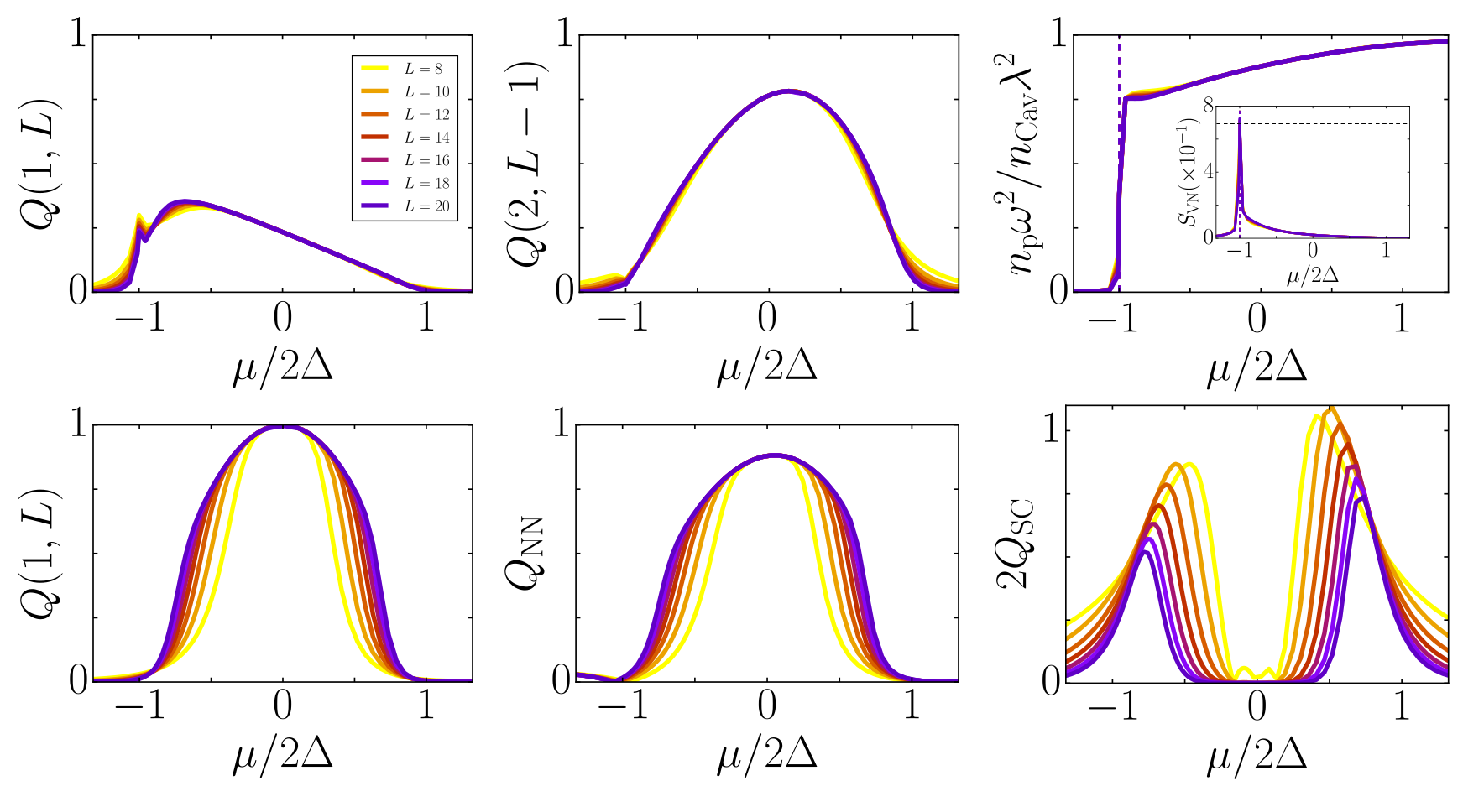}
\end{center}
\caption{Observables for different chain lengths. First row: edge geometry with $\lambda/2 \Delta=0.5$. Second row: bulk geometry with $\lambda/2 \Delta=0.67$. $Q_{\rm NN}$ denotes the correlations between the nearest neighbors of the cavity [i. e. $Q(L/2-1,L/2+2)$], and $Q_{\rm SC}$ are the correlations between the edges of the subchains [$Q(1,L/2-1)$]. For all the plots displayed $n_{\rm Cav}=2$, and $\omega/2\Delta=0.25$.}
  \label{fig:SM_SIZE}
\end{figure*}

\begin{figure*}[t!]
\begin{center}
 \includegraphics[width=1.0\textwidth]{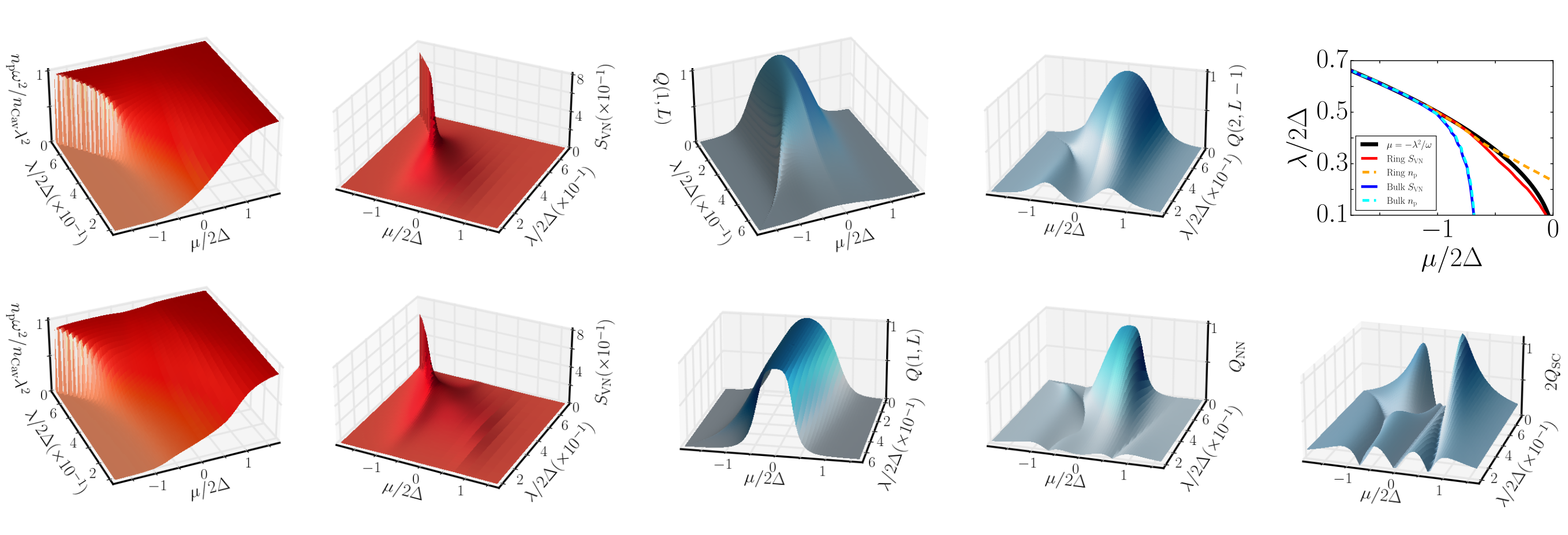}
\end{center}
\caption{Relevant observables as function of $\mu$ and $\lambda$. First row: edge geometry. Right corner: criticality comparison with different methodologies. The case $n_{\rm p}$ indicates that the curve is constructed from maximums in the first derivative of the number of photons for each $\lambda$. In contrast, $S_{\rm VN}$ shows the peaks in the Von Neumann entropy. For high $\lambda$, all criteria converge. Second row: bulk geometry. $Q_{\rm NN}$ denotes the correlations between the nearest neighbors of the cavity [i. e. $Q(L/2-1,L/2+2)$], and $Q_{\rm SC}$ are the correlations between the edges of the subchains [$Q(1,L/2-1)$]. For all the results displayed $n_{\rm Cav}=2$, $\omega/2\Delta=0.25$, and $L=8$.}
  \label{fig:SM_3D}
\end{figure*}

\begin{figure*}[t!]
\begin{center}
 \includegraphics[width=1.0\textwidth]{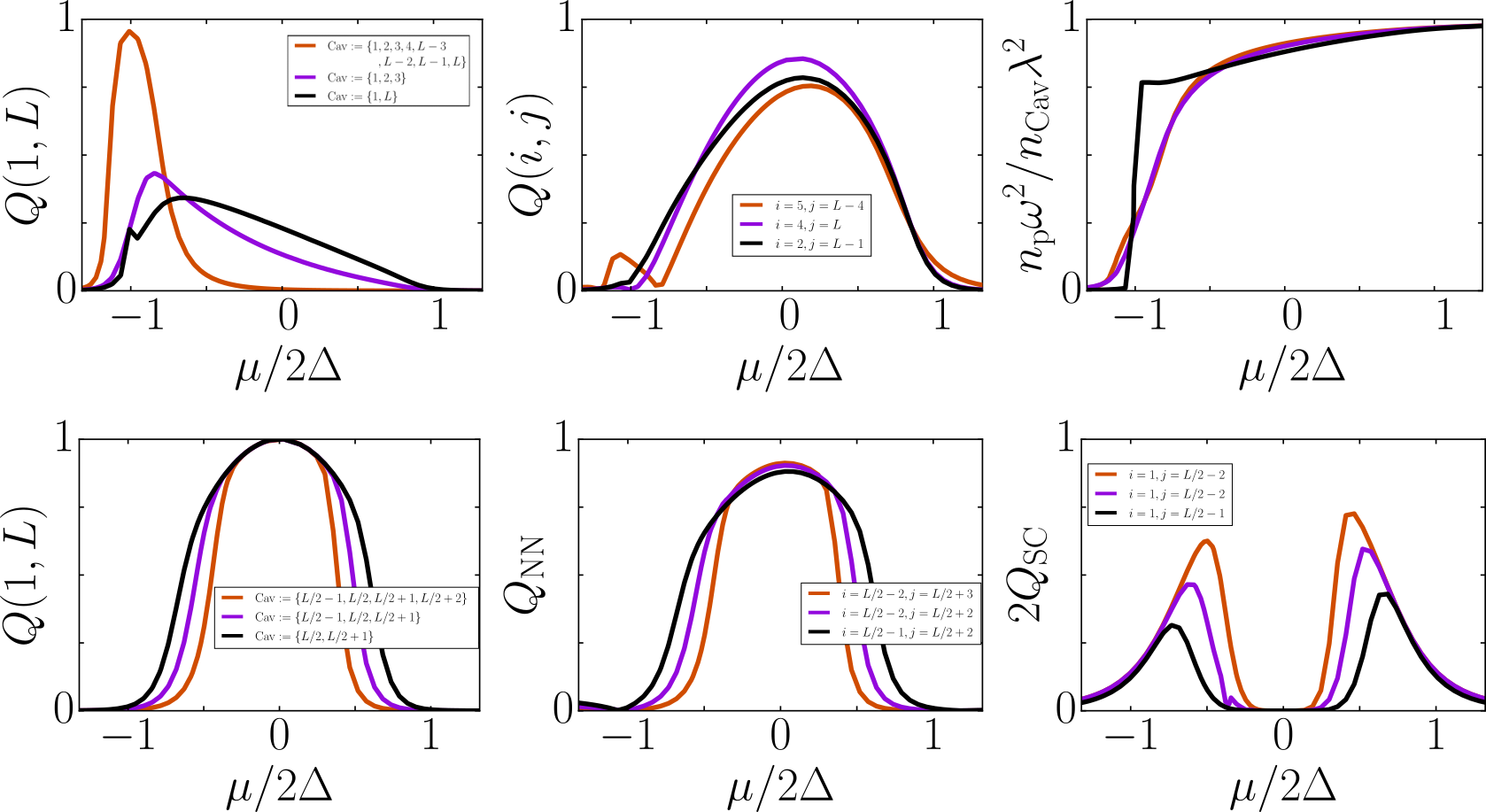}
\end{center}
\caption{Observables for different cavity configurations. First row: edge geometry with $\lambda/2 \Delta=0.5$. Second row: bulk geometry with $\lambda/2 \Delta=0.67$. $Q_{\rm NN}$ denotes the correlations between the nearest neighbors of the cavity, and $Q_{\rm SC}$ are the correlations between the edges of the subchains. The legends with the information of $i$ and $j$ provide the relevant indices for a given $\rm Cav$. For all the plots displayed $\omega/2\Delta=0.25$, and $L=16$.}
  \label{fig:SM_DifCavs}
\end{figure*}

In the present section, we compare our analytical description with DMRG simulations to show that the analysis accurately describes the physics of the system in a broad range of chain sizes and cavity geometries. Additionally, we complete the picture of the observables of interest.

First, in Fig. \ref{fig:SM_SIZE}, we show no qualitative change in the results by varying the system size. Quantitatively, the variations are small and do not change the physical meaning of our observations. Regarding the variations for the bulk geometry, note that the finite size effects for short chains are more dramatic than those presented in the MT. Still, the system has two pairs of indices defining the dominating correlations indicating the presence of the non-local qubits.

Next, Fig. \ref{fig:SM_3D} shows the behavior of the observables of interest as a function of $\lambda$ for the edge and bulk geometries displaying all the features mentioned in the MT. Again, for the bulk results, it is clear that the region of $\mu$ with dominating $Q(1,L)$ and $Q(L/2-1,L/2+2)$ shrinks in the topological phase in the DSC regime. However, the correlations $Q(1,L/2-1)$ and $Q(L/2+2,L)$ dominate in this region, and they complete the topological phase characteristics expected for $|\mu/2\Delta|<1$. From the entanglement results, it is clearly seen that the light-matter entanglement in the edge and bulk geometries displays a peak convergent to $\ln(2)$ for sufficiently high values of $\lambda$ though higher values are required to obtain the effect for the bulk geometry.

Finally, Fig. \ref{fig:SM_DifCavs} displays results for different cavity configurations. Although changing the number of sites involved in the interactions requires a higher value $\lambda$ to recover the abrupt transition, it is clear that we recover the same quantitative effect regarding the number of MZEMs and the localization of free Majorana fermions. This response occurs for different sizes and positions of the sites embedded in the cavity.

\begin{figure}[t!]
\begin{center}
 \includegraphics[width=1.0\linewidth]{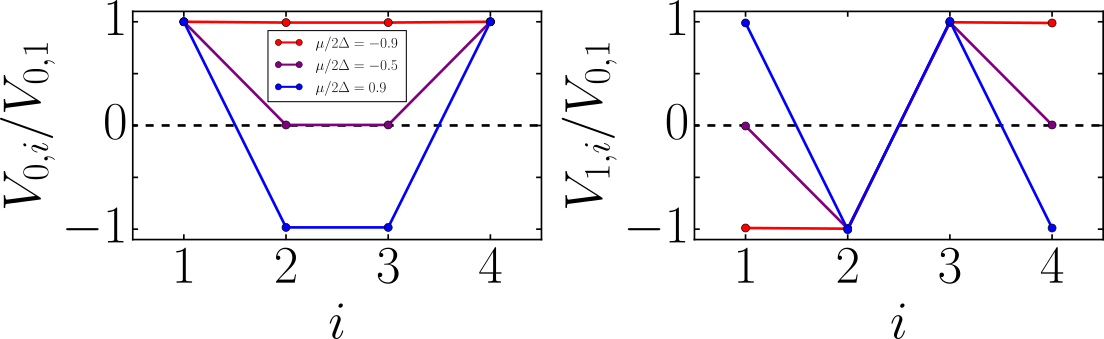}
\end{center}
\caption{Elements of the  vectors $V_0$ and $V_1$ normalized by the first element of $V_0$ for different values of $\mu$ in the topological phase. $i\in\{1,2,3,4\}$ is the index of the element of $V_0$ or $V_1$ For this example we used $L=16$, $n_{\rm Cav}=2$, $\omega/2\Delta=0.25$ and $\lambda/2\Delta=0.67$.}
  \label{fig:VS}
\end{figure}
\section{$Q(i,j)$ for the bulk geometry}
\label{sec:QforBulk}
In a Majorana chain, only MZEMs or nearly MZEMs are expected to be localized. Usually, they appear at the ends of a chain. The probability of finding a mode $k$ at the spatial site $i$ is $U_{k,i}^2+V_{k,i}^2$. For nearly MZEMs, the weights $U_{k,i}^2$ and $V_{k,i}^2$ are equivalent. Consequently, only nearly MZEMs lead the behavior of $Q(i,j)$ when $i$ and $j$ are chosen at the sites with dominant correlations. As mentioned in the MT, at the regime in which we obtain $\ket{\nu=n_{\rm Cav}}$, we identified those indices as $i,j \in \{1,L/2-1,L/2+2,L\}$. Moreover, due to the presence of the cavity, a new nearly MZEM emerges in the topological phase. Consequently, the degenerated $k$ modes are the main weights of the correlations. Let us define a 4-tuple $V_k=(V_{k,1},V_{k,L/2-1},V_{k,L/2+2},V_{k,L})$ and label the relevant modes describing free (and nearly free) Majorana fermions with $k=0$ and $1$. The value of the dominating $Q(i,j)$ will be dictated only by the elements of $V_k$. By numerical Bogoliubov diagonalization of the mean-field Hamiltonian in Eq. \eqref{eq:HChain_Effective} it can be observed that for values of $0.5<\mu/2\Delta< 1$ 
\begin{equation}
    V_0=(a,-a,-a,a); \quad 
    V_1=(a,-a,a,-a),
\end{equation}
with $a\in \mathbb{R}$ and $|a|<1$.
Thus, for these states $Q(1,L/2-2)$ and its symmetric correlation $Q(L/2+2,L)$ will go to $8a^2$. Conversely, $Q(1,L)$ and $Q(L/2-2,L/2+2)$ will approach 0. For the case $-1<\mu/2\Delta< -0.5$ we have
\begin{equation}
    V_0=(a,a,a,a); \quad 
    V_1=(-a,-a,a,a),
\end{equation}
which results in the same correlations as in the previous case.
On the other hand, when $0.5\leq|\mu/2\Delta|$ the $V_k$'s acquire the form
\begin{equation}
    V_0=(a,-b,b,-a); \quad 
    V_1=(b,-a,a,-b),
\end{equation}
with $a, b\in \mathbb{R}$ and $1>|a|\gg|b|$. From these results it follows that $Q(1,L)\approx Q(L/2-2,L/2+2) \approx 4a^2$ and $Q(1,L/2-2)=Q(L/2+2,L)\approx 0$. 

Fig. \ref{fig:VS} shows examples of the vectors $V_0$ and $V_1$ in the discussed regimes.

\section{First order perturbation theory}
\label{sec:firstOrderPerturbation}
Here, we address how to arrive at the equation
\begin{equation}
\label{Eq:FirstOrderPerturbationTheory}
    E_{\rm DSC}=E_{\rm MC}-n_{\rm Cav}\mu-\frac{n_{\rm Cav}\lambda^2}{\omega}
\end{equation}
 presented in MT. It was discussed in the MT that for the strong-coupling limit, the condition $\ket{\nu=n_{\rm Cav}}=N[\bigotimes_{j\in {\rm Cav}}\ket{1}_j]$ was required to minimize the energy of the GS. Here, $\ket{1}_j=\hat{c}_j^\dagger\ket{0}_j$ in the occupation basis and $N[\ket{\psi}]$ orders the string of creation operators contained in $\ket{\psi}$ to keep track of the signs coming from anti-commutation relations. The latter leads to the zeroth order energy $E_{n_{\rm Cav},0}=-\frac{n_{\rm Cav}\lambda^2}{\omega}$. Nevertheless, this restriction will not affect the sites outside the cavity. Therefore, the rest of the problem consists of finding the states of the sites $i\notin {\rm Cav}$ that minimize $\bra{\psi_0}\hat{H}_{\rm MC}\ket{\psi_0}$. Correspondingly, the latter minimization implies minimum $E_{\rm DSC}$. 

We are interested in the physics governing the abrupt change in the number of photons, which always occurs at $\mu<0$. Therefore, let us consider $\frac{\mu}{2\Delta}\ll-1$. In this limit, the GS of the isolated chain is expected to be characterized by $\ket{\psi_{\rm MC}^0}=N[\bigotimes_{i=1}^L\ket{0}_i]$. For this reason, the best suitable state for the sites at the rest of the chain will be $N[\bigotimes_{i\notin {\rm Cav}}\ket{0}_i]$ and therefore 
\begin{equation}
    \ket{\psi_0}=N[ \bigotimes_{i\notin {\rm Cav}}\ket{0}_i\bigotimes_{j\in {\rm Cav}}\ket{1}_j]\otimes\ket{-\lambda \sqrt{n_{\rm Cav}}/\omega}.
\end{equation} 
By inspecting $\hat{H}_{\rm MC}$ it is easy to see that
\begin{equation}
\label{eq:mulessthan-1}
    \bra{\psi_0}\hat{H}_{\rm MC}\ket{\psi_0}=E_{\rm MC}-n_{\rm Cav}\mu,
\end{equation}
with $E_{\rm MC}=\bra{\psi_{\rm MC}^0}\hat{H}_{\rm MC}\ket{\psi_{\rm MC}^0}$. Consequently, Eq. \ref{eq:mulessthan-1} immediately produces  Eq. \ref{Eq:FirstOrderPerturbationTheory}. This approximation provides good results even for regions in which $-1<\frac{\mu}{2\Delta}<0$. Following a similar procedure, analysis the value of $\bra{\psi_0}\hat{H}_{\rm MC}\ket{\psi_0}$ can be recovered considering the regimes $\frac{\mu}{2\Delta}\gg-1$ and $|\frac{\mu}{2\Delta}|\ll1$. The energy corrections under these considerations extend for finite values of $\mu$ (not shown). Note that the results do not depend on the position of the sites in the cavity $j$; they only rely on the size of the set $\rm Cav$. 
In the right upper corner of Fig. \ref{fig:SM_3D} it is shown that the transition predicted in Eq. \ref{Eq:FirstOrderPerturbationTheory} accurately describes the behavior for the edge and bulk geometries.

\end{document}